\documentstyle[12pt,aps,psfig]{revtex}
\begin{document}

\newcommand{\cd}{\makebox[0.08cm]{$\cdot$}}

\preprint{LBNL-52239}

\title{Rapidity Asymmetry in High-energy $d+A$ Collisions}

\author{Xin-Nian Wang}
\address{Nuclear Science Division, Mailstop 70R0319 \\
        Lawrence Berkeley National Laboratory, Berkeley, CA 94720}

\date{March 1, 2003}

\maketitle

\begin{abstract}
In contrast to the recent prediction of high $p_T$ hadron suppression
within the parton saturation model, it is shown that multiple 
parton scattering suffered by the projectile will 
enhance high $p_T$ hadron spectra in $d+A$ collisions
relative to a superposition of binary $p+p$ collisions 
at RHIC. A stronger enhancement in the forward rapidity 
region of the projectile is also predicted, resulting 
in a unique rapidity asymmetry of the hadron spectra at high $p_T$.
The shape of the rapidity asymmetry should be reversed 
for low $p_T$ hadrons that are dominated by soft and
coherent interactions which suppress hadron spectra in the 
projectile rapidity region. The phenomenon at the LHC 
energies is shown to be qualitatively different because 
of parton shadowing. 
\end{abstract}

\baselineskip=14pt

\section{Introduction}

Recent experiments \cite{phenix,star} at the Relativistic Heavy-ion 
Collider (RHIC) have shown a significant suppression of high $p_T$
hadron spectra in central $Au+Au$ collisions that was 
predicted \cite{gp90,wg92} as a consequence of parton energy 
loss or jet quenching in dense matter. In addition, the same 
mechanism is predicted to produce azimuthal anisotropy in high $p_T$
hadron spectra \cite{v2} that was also observed in experiments
at RHIC \cite{starv2}. This is a dramatic departure from the
heavy-ion collisions at the SPS energies where no significant
suppression of high $p_T$ spectra is observed \cite{wa98,wangsps}.
Since theoretical studies \cite{gw1,bdms,zhak,glv,wied} 
of parton propagation in a
dense medium all show that the parton energy loss induced by
multiple scattering is proportional to the gluon density,
RHIC data thus indicate an initial gluon density in central $Au+Au$ 
collisions at the RHIC energies that is much higher than that
in a large cold nucleus \cite{ww02}.

More precise extraction of the parton energy loss from the 
final hadron suppression in $A+A$ collisions, however, requires
the understanding of normal nuclear effects in $p+A$ collisions.
As pointed out by many early \cite{Lev,Ochiai,Wang98} and 
recent \cite{Zhang,Kopel,vg02} studies, 
the high $p_T$ hadron spectra can also be modified by 
initial multiple scatterings in $p+A$ and $A+A$ collisions giving 
rise to the observed Cronin effect. Within
a multiple scattering model, the high $p_T$ hadron spectra
are normally enhanced relative to $p+p$ collisions, except in the
kinematic region where the EMC effect \cite{emc} is important
(The EMC effect is the depletion of parton distributions
in $x\sim 0.2-0.8$ in nuclei caused by the nuclear binding effect).
Such a normal nuclear enhancement will not affect the
interpretation of the hadron suppression in central $A+A$
collisions as a consequence of jet quenching, though
inclusion of it is important for more precise extraction of
parton energy loss and the initial gluon density.

A very different mechanism was recently proposed for the observed
high $p_T$ hadron suppression based on the parton saturation 
model \cite{klm} which also predicts a similar suppression
in $p+A$ collisions at RHIC, in contrast to the predicted
enhancement by the multiple parton scattering model. While such a
model is not yet checked against the existing $p+A$ collisions
for energies up to $\sqrt{s}=40$ GeV where enhancement of hadron
spectra for $p_T>2$ GeV/$c$ has been successfully explained by the
multiple scattering model, the up-coming data of
$d+A$ collisions at RHIC will attest the relevance
of parton saturation at the RHIC energies.

In this letter, we will point out an additional feature
in the rapidity dependence of the Cronin enhancement due
to multiple parton scattering in $d+A$ collisions. Such a
rapidity dependence was also studied recently by Vitev \cite{Vitev:2003xu}.
However, we predict here a unique rapidity asymmetry of
high $p_T$ hadron spectra due to stronger Cronin
enhancement in the forward (projectile) region as a result
of the transverse momentum broadening of the initial partons
inside the projectile. The shape of the rapidity asymmetry will
also depend on the nuclear modification of the parton
distributions inside a nucleus, in particular at small $x$.
As one decreases $p_T$, the parton shadowing will reduce
the hadron spectra in the forward region, thus changing the
rapidity asymmetry. When soft and coherent interactions become
dominant at very low $p_T<1$ GeV/$c$, the shape of the rapidity
asymmetry will be reversed because of the strong 
suppression of hadron production in the projectile region
relative to a superposition of binary $p+p$ collisions.
We will calculate the rapidity asymmetry within a perturbative 
QCD (pQCD) parton model and study the effect of nuclear 
modification of parton distributions.

\begin{figure}
\centerline{\psfig{file=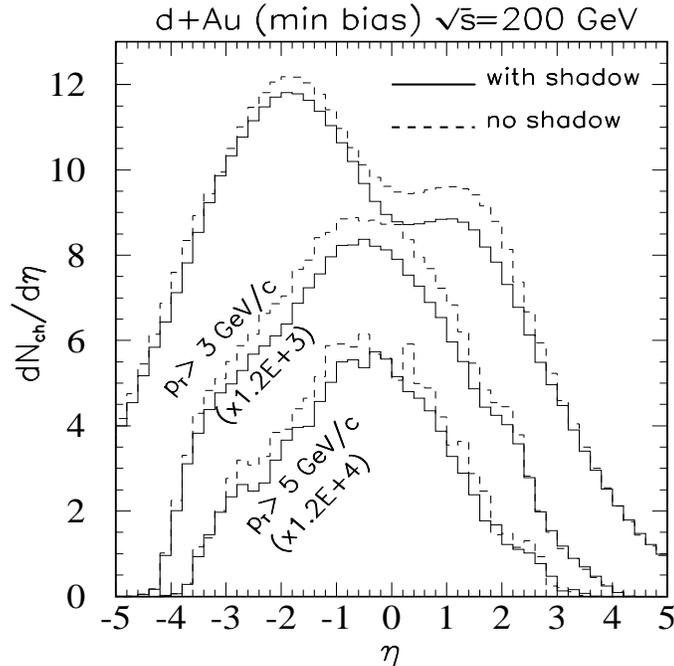,width=3.5in,height=3.5in}}
\caption{Pseudo-rapidity distribution of charged hadrons in 
minimum-biased $d+Au$ collisions at $\sqrt{s}=200$ GeV from 
the HIJING \protect\cite{hijing} model.}
\label{fig:hijing}
\end{figure}

In a Glauber multiple parton scattering model, large $p_T$ 
spectra are generally enhanced relative to the binary model 
of hard scattering. It can be shown \cite{ww01}
that a combination of absorptive corrections and the 
power-law behavior of perturbative parton cross section 
leads to a nuclear enhancement at high $p_T$ that decreases 
as $1/p_T^2$. The same absorptive processes suppress 
the spectra relative to the binary model at low $p_T$, 
where soft processes dominate and the $p_T$ spectra deviate 
from a power-law behavior. Since the pQCD model cannot be applied
to soft processes, one has to resort to phenomenological models
like the string model in which coherent particle production is
modeled by a string excitation for each participant nucleon.
Shown in Fig.~\ref{fig:hijing} is the rapidity distribution
(integrated over transverse momentum) of charged hadrons
in minimum-biased $d+Au$ collisions at $\sqrt{s}=200$ GeV
from the HIJING Monte Carlo model \cite{hijing}, which employs
the string model for soft and coherent interactions.
Because most of the soft particles are produced through
string fragmentation, their number should then be proportional
to the number of participants. The rapidity distributions of
hadrons from the string should also follow their parent nucleons.
Since there are more target nucleon participants than projectile
nucleons in $d+A$ collisions, the hadron rapidity distribution is thus
asymmetrical with respect to $\eta=0$. For high $p_T$ hadrons,
the underlying processes are hard parton scatterings. The
multiplicity should be approximately proportional to the
number of binary scatterings and the rapidity distributions
should be approximately symmetric with respect to $\eta=0$.
This is roughly the case for high $p_T$ hadrons in HIJING
model as shown in Fig.~\ref{fig:hijing}.
One, however, can notice some asymmetric effect in the
rapidity distribution of large $p_T$ hadrons. This is partially
due to the coherence between transverse jets and the beam remnants,
which is responsible for the asymmetric pedestal effect underlying
a jet event in $p+A$ collisions. In addition, energy and quark
number conservation will also cause some asymmetric effects.
These effects are most important in the large rapidity region.
It is expected that they are small in the central rapidity region.

Though the HIJING model has incorporated hard processes, it has not
included transverse momentum broadening of initial partons. For
this purpose, we use the pQCD model as employed in Ref.~\cite{Wang98}.
We will use a lowest order (LO) pQCD-inspired parton model in which
the inclusive particle production cross section in $pp$ collisions
is given by \cite{Owens}
\begin{eqnarray}
  \frac{d\sigma^h_{pp}}{dyd^2p_T}&=&K\sum_{abcd}
  \int dx_a dx_b d^2k_{aT} d^2k_{bT} g_p(k_{aT},Q^2) g_p(k_{bT},Q^2) 
  \nonumber \\ & & f_{a/p}(x_a,Q^2)f_{b/p}(x_b,Q^2) 
  \frac{D^0_{h/c}(z_c,Q^2)}{\pi z_c}
  \frac{d\sigma}{d\hat{t}}(ab\rightarrow cd), \label{eq:nch_pp}
\end{eqnarray}
where $D^0_{h/c}(z_c,Q^2)$ is the fragmentation function of 
parton $c$ into hadron $h$ as parameterized in Ref.~\cite{bkk} 
from $e^+e^-$ data, and $z_c$ is the momentum fraction of a parton 
jet carried by a produced hadron. The $K\approx 1.5$ 
(at $\sqrt{s}=200$ GeV) factor is used to account for higher 
order QCD corrections to the jet production cross 
section. The parton distributions $f_{a/N}(x,Q^2)$ 
in a nucleon are given by the MRS D$-^{\prime}$ parameterization \cite{mrs}.
The initial transverse momentum distribution $g_N(k_T,Q^2)$ is 
assumed to have a Gaussian form,
\begin{equation}
    g_N(k_T,Q^2)=\frac{1}{\pi \langle k^2_T\rangle_N} 
    e^{-k^2_T/\langle k^2_T\rangle_N}.
\end{equation}
Following \cite{Owens2}, we choose a $Q$-dependent average initial
transverse momentum,
\begin{equation}
\langle k^2_T\rangle_N(Q^2)= 1.2 ({\rm GeV}^2) + 0.2\alpha_s(Q^2) Q^2,
\label{kperp}
\end{equation}
which should include both the intrinsic and pQCD radiation-generated 
transverse momentum in this LO calculation.
The form of the $Q$-dependence and the parameters are chosen 
to reproduce the experimental data \cite{Wang98}, especially at
low energies. Following the same approach as in 
Refs. \cite{Owens2,Feynman}, we choose $Q^2$ to 
be $Q^2=2\hat{s}\hat{t}\hat{u}/(\hat{s}^2+\hat{t}^2+\hat{u}^2)$.

To take account of multiple initial-state scattering, 
we assume that the inclusive differential
cross section for large $p_T$ particle production is still 
given by a single hard parton-parton scattering. 
However, due to multiple parton scattering prior to 
the hard processes, we consider the initial
transverse momentum $k_T$ of the beam partons to be broadened. Assuming
that each scattering provide a $k_T$ kick which also has a Gaussian
distribution, we can effectively change the width of the initial
$k_T$ distribution. Then the single inclusive particle cross section
in minimum-biased $p+A$ collisions is,
\begin{eqnarray}
  \frac{d\sigma^h_{pA}}{dyd^2p_T}&=&K\sum_{abcd} \int d^2b t_A(b)
  \int dx_a dx_b d^2k_{aT} d^2k_{bT}  g_A(k_{aT},Q^2,b) g_p(k_{bT},Q^2)
  \nonumber \\ & & 
  f_{a/p}(x_a,Q^2)f_{b/A}(x_b,Q^2,b)
  \frac{D^0_{h/c}(z_c,Q^2)}{\pi z_c}
  \frac{d\sigma}{d\hat{t}}(ab\rightarrow cd), \label{eq:nch_pA}
\end{eqnarray} 
where $t_A(b)$ is the nuclear thickness function normalized 
to $\int d^2b t_A(b)=A$.  We will use the Woods-Saxon form of 
nuclear distribution for $t_A(b)$ throughout this paper 
unless specified otherwise. The parton distribution per 
nucleon inside the nucleus (with atomic mass number
$A$ and charge number $Z$) at an impact parameter $b$,
\begin{equation}
    f_{a/A}(x,Q^2,b)=S_{a/A}(x,b)\left[ \frac{Z}{A}f_{a/p}(x,Q^2)
    +(1-\frac{Z}{A}) f_{a/n}(x,Q^2)\right], \label{eq:shd}
\end{equation}
is assumed to be factorizable into the parton distribution in a
nucleon $f_{a/N}(x,Q^2)$ and the nuclear modification factor
$S_{a/A}(x,b)$, for which we take both the new HIJING \cite{LW} 
and EKS \cite{eks} parameterizations. The initial parton 
transverse momentum distribution inside a projectile nucleon 
going through the target nucleus at an impact parameter $b$ is
still a Gaussian with a broadened width
\begin{equation}
\langle k^2_T\rangle_A(Q^2)=\langle k^2_T\rangle_N(Q^2)
    +\delta^2(Q^2)(\nu_A(b) -1).
\end{equation}
The broadening is assumed to be proportional to the number of
scattering $\nu_A(b)$ the projectile suffers inside the nucleus,
which is assumed to given by
\begin{equation}
    \nu_A(b)=\sigma_{NN} t_A(b)=
    \sigma_{NN} \frac{3 A}{2\pi R_A^2}\sqrt{1-b^2/R_A^2}
\end{equation}
in a hard sphere nuclear distribution, where $R_A=1.12 A^{1/3}$ fm 
and $\sigma_{NN}$ is the inelastic nucleon-nucleon cross section.
We also assume that $k_T$ broadening during each 
nucleon-nucleon collision $\delta^2$ also depends 
on the hard momentum scale $Q=P^{\rm jet}_T$ in the
parameterized form,
\begin{equation}
\delta^2(Q^2)=0.225\frac{\ln^2(Q/{\rm GeV})}{1+\ln(Q/{\rm GeV})}
 \;\;\;{\rm GeV^2}/c^2,
\end{equation}
which is chosen to best fit the existing experimental data in $p+A$
collisions \cite{Wang98} up to $\sqrt{s}=40$ GeV. The predictive 
power of this model lies in the energy and flavor dependence of 
the hadron spectra. It is straightforward to also calculate
hadron spectra in  $d+A$ collisions, incorporating the 
initial $k_T$ broadening and parton shadowing. We have tried
different forms of nuclear distribution for deuteron and find
little difference in the final results. So we will still use the
Woods-Saxon distribution for $t_d(b)$. In heavy nuclear $A+A$
collisions, one can incorporate parton energy loss induced by the
dense medium through modified fragmentation functions \cite{wh}.

The pQCD model described above has been compared to
experimental data for $pp$, $p\bar{p}$ and $pA$ collisions 
at various energies \cite{Wang98}. One found that both the 
intrinsic $k_T$ and the nuclear broadening are very important 
to describe the existing data in $pp$ and $pA$ collisions, 
especially at around SPS energies. One can find some more 
detailed description of this pQCD-inspired  model in 
Ref. \cite{Wang98}. We will restrict ourselves to the study of the
rapidity dependence of the nuclear modification in $d+A$ 
collisions at RHIC and LHC energies in this paper. 

\begin{figure}
\centerline{\psfig{file=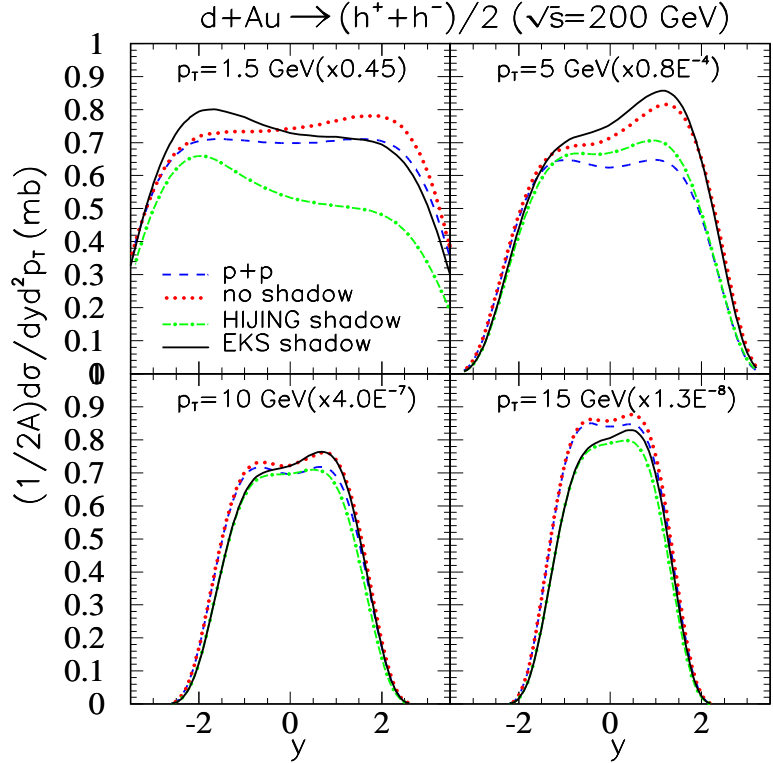,width=4.0in,height=4.0in}}
\caption{Calculated rapidity distributions of charged hadrons with 
large transverse momentum in minimum-biased $d+Au$ collisions 
at $\sqrt{s}=200$ GeV. The dashed lines correspond to a superposition
of binary $N+N$ collisions. The dotted lines have nuclear broadening
of initial parton transverse momentum but without parton shadowing. The
solid and dot-dashed lines use the EKS \protect\cite{eks} and
HIJING \protect\cite{LW} parameterization of parton shadowing,
respectively. The spectra have been scaled by the numbers in
parentheses.}
\label{fig:dy200}
\end{figure}

Neglecting initial transverse momentum, the initial momentum
fractions are related to the transverse momentum and
rapidities of the final jets by $x_1=x_T(e^{y_1}+e^{y_2})/2$,
$x_2=xT(e^{-y_1}+e^{-y_2})/2$, $x_T=2E_T/\sqrt{s}$, where
$E_T$, $y_1$ and $y_2$ are the transverse momentum and
rapidities of the produced jets, respectively. Large positive
rapidities, therefore, correspond to large parton fractional
momentum $x_1$ from the projectile and small momentum fraction
$x_2$ from the target. Conversely, negative rapidities correspond
to small $x_1$ and large $x_2$. In our parton model calculation, 
we assume the final hadron rapidity to be the same as that of the 
fragmenting jets. Shown in Fig.~\ref{fig:dy200} are the 
rapidity distributions of the charged hadron spectra for four
different values of $p_T$. Without nuclear modification of the
parton distributions (shown as dotted lines), the $k_T$ broadening 
of the projectile partons enhances the particle spectra. The 
enhancement is the strongest in the forward (projectile) region,
thus giving rise to a rapidity asymmetry. 
The EKS \cite{eks} parameterization has a strong anti-shadowing
at $x_2\sim 0.1-0.2$ for partons from the nuclear target. Such
strong anti-shadowing further enhances the rapidity 
asymmetry (shown as solid lines). However, the HIJING \cite{LW} 
parameterization has mostly shadowing in this region and thus
reduces the rapidity asymmetry in the spectra (dot-dashed line)
which is still visible for $p_T>3$ GeV/$c$. 
At $p_T>10$ GeV/$c$, the target parton distribution 
at $x_2\sim 0.2 - 0.8$ is suppressed due to the nuclear
binding, which is known as the EMC effect \cite{emc}. 
This reduces the hadron spectra in the backward (target) 
rapidity region and thus further enhances the rapidity asymmetry.

As one decreases the transverse momentum so that parton shadowing
in the target becomes significant, the hadron spectra in the
forward region are strongly suppressed. It can even overcome the
enhancement caused by transverse momentum broadening. In this case
the rapidity asymmetry is reversed, as shown in Fig.~\ref{fig:dy200}
for $p_T=1.5$ GeV/$c$. This is much like the asymmetric rapidity
distribution of soft particles produced via soft and coherent
interactions as shown in Fig.~\ref{fig:hijing}. The validity of
the parton model at $p_T=1.5$ GeV/$c$ may be questionable. 
However, it clearly shows the trend of the rapidity
asymmetry as one changes the value of $p_T$.

\begin{figure}
\centerline{\psfig{file=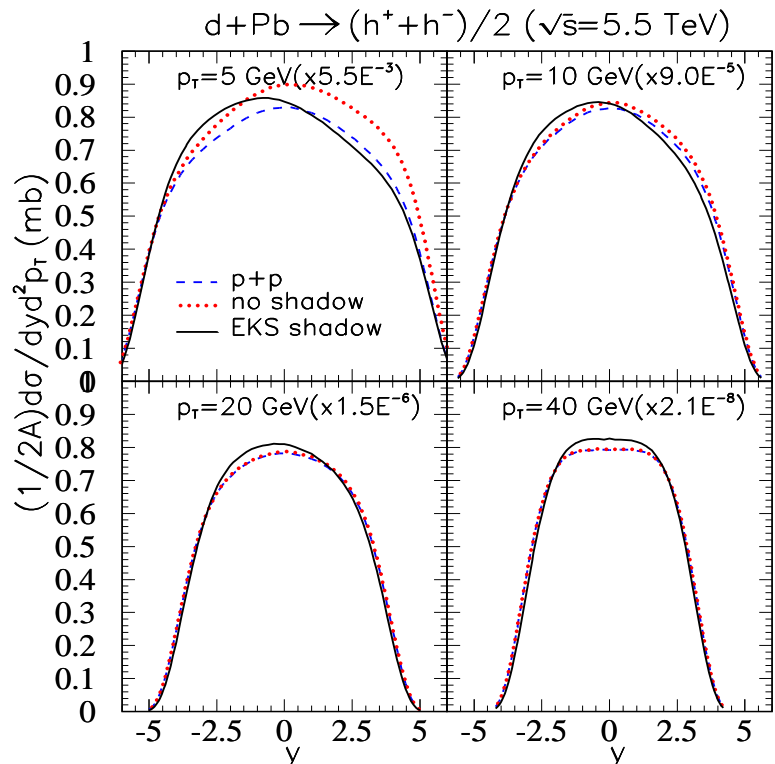,width=4.0in,height=4.0in}}
\caption{Calculated rapidity distributions of charged hadrons with 
large transverse momentum in minimum-biased $d+Pb$ collisions 
at $\sqrt{s}=5.5$ TeV.}
\label{fig:dy5500}
\end{figure}

We also show in Fig.~\ref{fig:dy5500} similar rapidity
distributions for $d+Pb$ collisions at $\sqrt{s}=5.5$ TeV where
one has access to much larger values of transverse momentum.
At such high energies and transverse momenta, 
the parton scattering cross is much
flatter in $p_T$ than at lower energies. Therefore, the
final hadron spectra are less sensitive to the transverse
momentum broadening due to initial multiple scatterings.
The rapidity asymmetry caused by the $k_T$ broadening is
thus very weak, as shown by the dotted lines. The nuclear
modification of the parton distributions in the target
is causing most of the rapidity asymmetry in the spectra
at these energies. Here we used only EKS \cite{eks}
parameterization of nuclear modification of the parton
distributions. The HIJING \cite{LW} parameterization does not
have any scale dependence, which is very important at LHC
energies since the shadowing still has significant effects on
jet production with large transverse momentum.

\begin{figure}
\centerline{\psfig{file=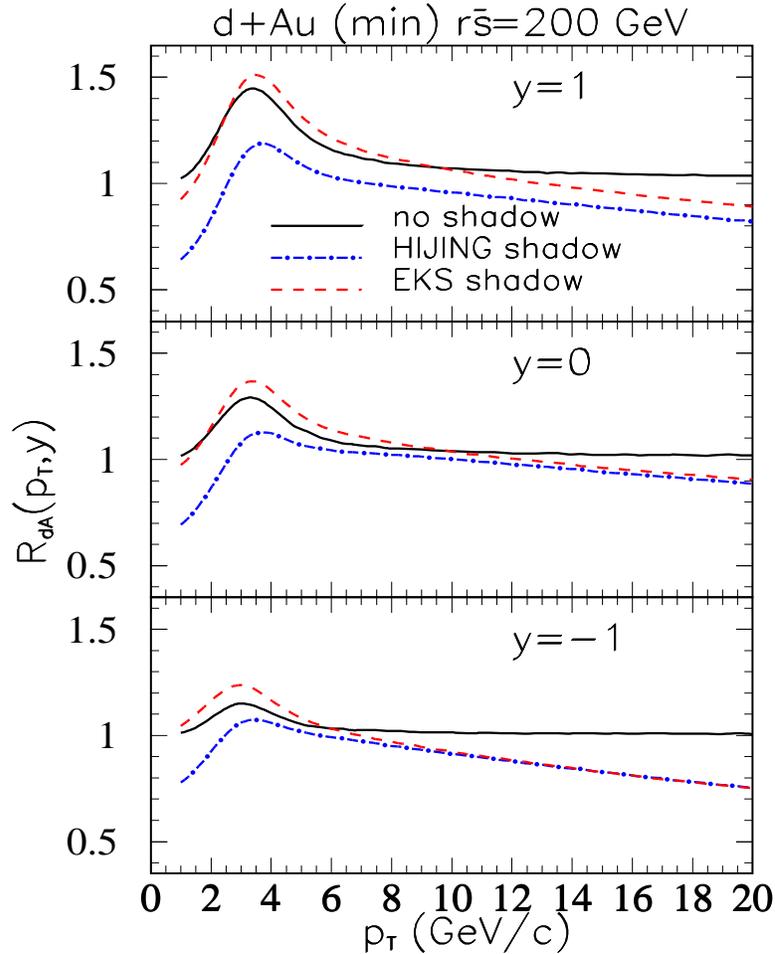,width=4.0in,height=5.0in}}
\caption{The nuclear modification factor $R_{dA}(p_T,y)$ for $d+Au$
   collisions at $\sqrt{s}=200$ GeV. $k_T$ broadening and different
  parameterization of parton shadowing are included.}
\label{fig:r200}
\end{figure}

\begin{figure}
\centerline{\psfig{file=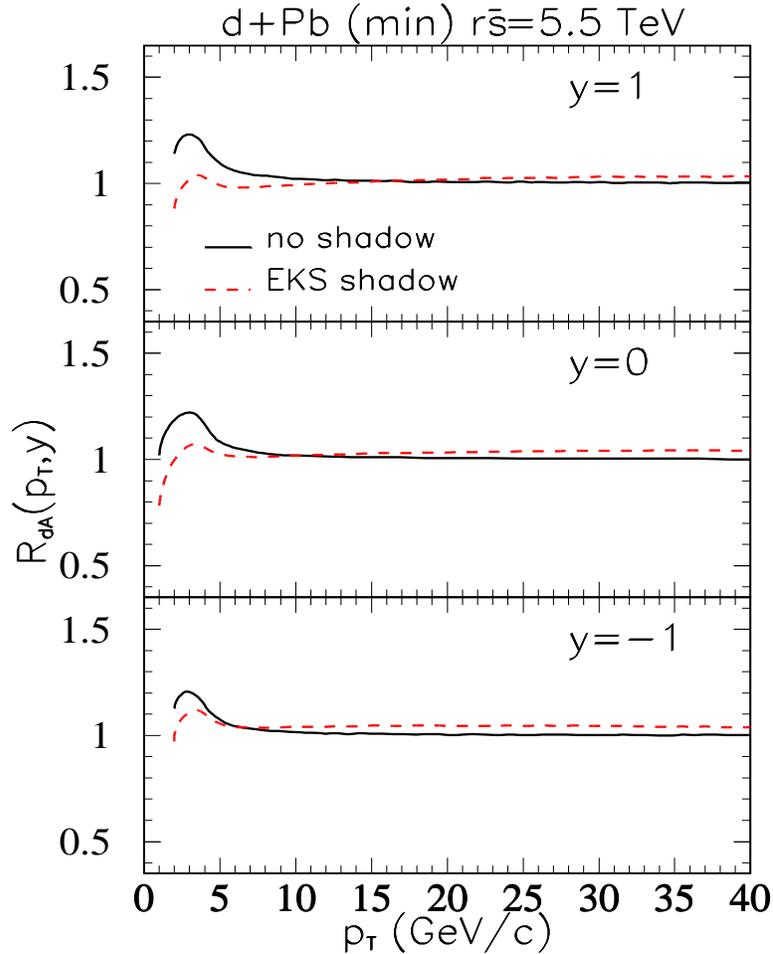,width=4.0in,height=5.0in}}
\caption{The nuclear modification factor $R_{dA}(p_T,y)$ for $d+Pb$
  collisions at $\sqrt{s}=5.5$ TeV.}
\label{fig:r5500}
\end{figure}

To demonstrate the $p_T$ dependence of the effect of parton 
shadowing and $k_T$ broadening, we show in Fig.~\ref{fig:r200} 
and \ref{fig:r5500} the nuclear modification factors 
defined as the ratio of charged hadron spectra in $d+A$ 
over that in $p+p$ normalized to the averaged number of
binary nucleon collisions,
\begin{equation}
  R_{dA}(p_T,y)\equiv \frac{d\sigma^h_{AB}/dyd^2p_T}
  {2A d\sigma^h_{pp}/dyd^2p_T}. \label{eq:ratio}
\end{equation}
At the RHIC energy, the transverse momentum broadening enhances
the hadron spectra for $p_T=2-8$ GeV/$c$ in both forward and
backward rapidity regions. The exact enhancement in this
$p_T$ region depends on the parton shadowing and anti-shadowing.
The enhancement in the forward rapidity is larger than in the
backward region as demonstrated by the rapidity asymmetry
in Fig.~\ref{fig:dy200}. For $p_T>8$ GeV/$c$, the EMC
effect of the nuclear modification of parton distribution
starts to suppress the hadron spectra. Such a suppression
is stronger in the target rapidity region than in the
projectile region. We also find that the suppression due to the
EMC effect is stronger for kaons than pions. 
For $p_T<2$ GeV/$c$, parton shadowing 
suppresses the hadron spectra in the case of HIJING 
parameterization. However, the pQCD model might not be valid
anymore in this small $p_T$ region for a quantitative calculation.
The nuclear modification factor
for $d+Pb$ collisions at the LHC energy shown 
in Fig. ~\ref{fig:r5500} has much smaller variation and
the Cronin enhancement is also smaller.

In summary, we have studied the rapidity distribution of hadron
spectra in high-energy $d+A$ collisions, in particular the
rapidity asymmetry caused by multiple parton scatterings. 
The effects of initial multiple parton 
scatterings are incorporated via an impact-parameter dependent 
nuclear shadowing of parton distributions and the broadening of 
the initial $k_T$ carried by partons before they collide and 
produce high $p_T$ hadrons. At low $p_T$ parton shadowing
suppresses the hadron spectra in the projectile rapidity
region, giving rise to a rapidity asymmetry much like the
soft particle production via soft and coherent interactions
in a string model. However, as one increases $p_T$, hadron
production is dominated by hard parton scatterings and
the rapidity distribution is becoming more symmetric.
Within a pQCD model, transverse momentum broadening via
initial multiple scattering enhances the hadron spectra in
the projectile region at moderate $p_T=3-8$ GeV/$c$, causing 
a rapidity asymmetry opposite to that of soft hadrons. 
This is in sharp contrast to the prediction of the parton
saturation model \cite{klm}. The coming data of $d+Au$ collisions from RHIC
can easily distinguish those two models and will be important
to verify whether jet quenching due to parton energy loss
is truly the underlying mechanism for the observed hadron
suppression in central $A+A$ collisions.

We should also caution that our pQCD model cannot take into
account the non-perturbative effects that can cause some
rapidity asymmetry at very large rapidities. Such non-perturbative
effect is responsible for the asymmetric pedestal effect underlying
a jet event in $p+A$ collisions. Our conclusions are more
robust within the central rapidity region where such non-perturbative
effect and constraints by total energy conservation are not yet important.

This work was supported by the the Director, Office of Energy
Research, Office of High Energy and Nuclear Physics, Division of
Nuclear Physics, and by the Office of Basic Energy Science,
Division of Nuclear Science, of  the U.S. Department of Energy
under Contract No. DE-AC03-76SF00098.

\end{document}